\documentclass[twocolumn]{aa}

\def\beq{\begin{equation}}
\def\enq{\end{equation}}
\def\beqn{\begin{eqnarray}}
\def\enqn{\end{eqnarray}}

\begin{document}
\title{Could electron-positron annihilation lines in the Galactic
center result from pulsar winds?}

\author{W. Wang\inst{1,3}, C.S.J. Pun\inst{2} and K. S. Cheng\inst{2}}
\offprints{W. Wang}

\institute{Max-Planck-Institut f\"ur extraterrestrische Physik,
Postfach 1312, 85741 Garching, Germany \\
\email{wwang@mpe.mpg.de} \and Department of Physics, University of
Hong Kong, Pokfulam Road, Hong Kong, China \\
\and National Astronomical Observatories, Chinese Academy of
Sciences, Beijing 100012, China}

\date{August 24, 2005}

\abstract{ The observations of a strong and extended
positron-electron annihilation line emission in the Galactic
center (GC) region by the Spectrometer on the International
Gamma-Ray Astrophysical Laboratory (SPI/INTEGRAL) are challenging
to the present models of positron sources in the Galaxy. In this
paper, we study the possibility of pulsar winds in the GC to
produce the 511 keV line. We propose that there may exist three
possible scenarios of pulsar winds as the positron sources: normal
pulsars; rapidly spinning strongly magnetized neutron stars
(magnetars) in gamma-ray burst (GRB) progenitors; a population of
millisecond pulsars in the Galactic center region. These $e^\pm$
pairs could be trapped in the region by the magnetic field in the
Galactic center, and cool through the synchrotron radiation and
Coulomb interactions with the medium, becoming non-relativistic
particles. The cooling timescales are lower than the diffuse
timescale of positrons, so low energy positrons could annihilate
directly with electrons into 511 keV photons or form positronium
before annihilation. We find that normal pulsars cannot be a
significant contributor to the positron sources. Although
magnetars in the GC could be potential sources of positrons, their
birth rate and birth locations may impose some problems for this
scenario. We believe that the most likely candidate positron
sources in the GC may be a population of millisecond pulsars in
the GC. Our preliminary estimations predict the e$^\pm$
annihilation rate in the GC is $\geq 5\times 10^{42}$ s$^{-1}$
which is consistent with the present observational constraints.
Therefore, the $e^\pm$ pairs from pulsars winds can contribute
significantly to the positron sources in the Galactic center
region. Furthermore, since the diffusion length of positrons is
short, we predict that the intensity distribution of the
annihilation line should follow the distribution of millisecond
pulsars, which should correlate to the mass distribution in the
GC.

\keywords{Galaxy: center -- Gamma-rays: theory -- pulsars:
general }}

\authorrunning{W. Wang et al.}
\titlerunning{Electron-positron annihilation lines by pulsar winds?}

\maketitle

\section{Introduction}

Since the first detection (Johnson \& Haymes 1973) and the
subsequent identification (Leventhal et al. 1978) of the Galactic
511 keV annihilation line, the origins of the galactic positrons
have become a lively topic of scientific debate. With the launch
of INTEGRAL gamma-ray observatory in 2002, the SPI telescope on
board provides a strong constraints on the morphology and
intensity of the 511 keV line emission from the Galactic center
(Kn\"odlseder et al. 2003; Jean et al. 2003). The data analyses
show the line emitting source is diffuse, and the line flux within
5$^\circ$ of the GC amounts to $\sim (9.9\pm 4) \times 10^{-4}{\rm
photon\ cm^{-2}\ s^{-1}}$ (Kn\"odlseder et al. 2003),
corresponding to a luminosity of $\sim 10^{36}{\rm erg\ s^{-1}}$.
The high line luminosity requires that the positron injection rate
in the GC should be around $(3-6)\times 10^{42}\ e^+{\rm s^{-1}}$.
Recently, the analyses of the observational data by SPI/INTEGRAL
with deep Galactic center region exposure (longer than 4 Ms) show
that the spatial distribution of 511 keV line appears centered on
the Galactic center (bulge component), with no contribution from a
disk component (Teegarden et al. 2005; Kn\"odlseder et al. 2005;
Churazov et al. 2005). Churazov et al. (2005)'s analysis suggested
that the positron injection rate is up to $10^{43}\ e^+{\rm
s^{-1}}$ within $\sim 6^\circ$, and an intrinsic annihilation line
width $\sim 2.37\pm 0.25$ keV.

The potential positron sources include neutron stars or black
holes (Lingenfelter \& Ramaty 1983), $^{56}$Co $\beta$-decays in
supernova remnants (Lingenfelter \& Ramaty 1979; Ellison, Frank \&
Ramaty 1990), other radioactive nuclei formed by nucleosynthesis
in supernova, nova, red giants, and Wolf-Rayet stars (Ramaty et
al. 1979), cosmic ray interactions with the interstellar medium
(Kozlovsky et al.1987), pair production by gamma-ray photons
interacting with starlight photons in the interstellar medium
(Mastichiadis, Protheroe \& Stephen 1991), and gamma-ray cascades
at the polar caps of pulsars (Harding \& Ramaty 1987; Boulares
1989), electron-positron pairs produced by the pulsar winds (Chi,
Cheng \& Young 1996), and probably gamma-ray bursts (GRB,
Lingenfelter \& Hueter 1984). However, the recent results from
SPI/INTEGRAL on the 511 keV annihilation line emission in the
Galactic center show a diffuse source with a high line luminosity,
and present a challenge to the present models of the origin of the
galactic positrons. Hence, Cass\'e et al. (2004) argued that
normal supernova explosion in the GC cannot contribute to the
required positrons, but hypernovae (Type Ic supernovae/gamma-ray
bursts) in the Galactic center may be the possible sources of
galactic positrons. In addition, annihilations of light dark
matter particles into $e^\pm$ pairs (Boehm et al. 2004) have been
also suggested to be the potential origin of the 511 keV line in
the GC.

In this paper, we will study the possible contribution to the
positrons in the Galactic center from pulsar winds. We suggest
that there are three possibilities of pulsar winds as positron
sources in the Galactic center: normal pulsars (e.g. Crab and Vela
like pulsars); the rapidly spinning strongly magnetized neutron
stars which are possible GRB progenitors (Usov 1992); a
millisecond pulsar population (Wang, Jiang \& Cheng 2005). In \S
2, we will first study the possibility that electron-positron
pairs can be produced by the pulsar winds. In \S 3, we discuss
electron-positrons produced in three possible scenarios in detail.
These $e^\pm$ pairs could be trapped in the Galactic center by the
magnetic field. In \S 4, we find that the diffuse timescale of
electron-positrons in the GC is much longer than the synchrotron
cooling timescale and energy loss timescale by Coulomb
interactions in the medium, so that $e^\pm$ pairs will become
non-relativistic, and then annihilate into 511 keV emission lines.
The summary and discussions will be shown in \S 5.

\section{Electron-positron winds from pulsars}
It is well known that relativistic particles from pulsar winds
interacting with the interstellar medium form the synchrotron wind
nebulae (see the reviews by Arons 1998; Kaspi, Roberts \& Harding
2004). Chi, Cheng \& Young (1996) have proposed that the bulk of
the comic positrons could originate from pulsar winds. In this
paper, we will consider electron-positron pair production which
occurs in the pulsar outer-magnetospheric region, the so-called
outer gap (Cheng, Ho \& Ruderman 1986).

Zhang \& Cheng (1997) have developed a self-consistent mechanism
to describe the high energy radiation from spin-powered pulsars.
In the model, relativistic charged particles from a thick outer
magnetospheric accelerator (outer gap) radiate through the
synchro-curvature radiation mechanism (Cheng \& Zhang 1996) rather
than the synchrotron and curvature mechanisms in general,
producing non-thermal photons from the primary $e^\pm$ pairs along
the curved magnetic field lines in the outer gap. The criterion
for the existence of the outer gap requires the fractional size of
the outer gap $f<1$, which is the ratio between the mean vertical
separation of the outer gap boundaries in the plane of the
rotation axis and the magnetic axis to the light cylinder radius.
$f$ is limited by the pair production between the soft thermal
X-rays from the neutron star surface and the high energy gamma-ray
photons emitted from the outer gap region, and can be approximated
as $f\simeq 5.5 P^{26/21}B_{12}^{-4/7}$ (Zhang \& Cheng 1997),
where $P$ is the spin period, and $B_{12}$ is the surface magnetic
field in units of $10^{12}$ G.

The pair production mechanism is a synchrotron photon cascade in a
strong magnetic field. According to Halpern \& Ruderman (1993),
each primary electron-positron from an outer gap will have an
energy \beq E_p=\gamma_pm_ec^2=5.7\times 10^{12}P^{1/3} {\rm eV}
\enq before it strikes the neutron star surface, where $\gamma_p$
is the Lorentz factor of the primary electrons, $m_e$ is the rest
mass of the electron and $c$ is the light speed. These primary
electron-positrons will continue to emit curvature photons with a
typical energy \beq E_{\rm cur}={3\over 2}\gamma_p^3 {c\over s}
\hbar =6.4\times 10^8 P^{1/2}{\rm eV}, \enq where $\hbar$ is the
Planck constant, $s=(RR_L)^{1/2}$ is the curvature radius,
$R=10^6$ cm is the radius of the neutron star, and $R_L=cP/2\pi$
is the light cylinder distance. These curvature photons will be
converted into secondary $e^\pm$ pairs in the local magnetic field
when their energy satisfies (Ruderman \& Sutherland 1975) \beq
E\geq E_{\rm crit}\equiv {2m_ec^2\over 15}{B_q\over
B_s}=3B_{s,9}^{-1} {\rm GeV}, \enq where $B_q=4.4\times 10^{13}$ G
is a critical magnetic field, $B_{s,9}$ is the local surface
magnetic field of the neutron star in units of $10^9$ G.

It has been proposed that there is strong multipole magnetic field
near the stellar surface although a global dipole magnetic field
gives a good description of the magnetic field far from the star
(Ruderman \& Sutherland 1975; Ruderman 1991; Arons 1993). The
typical radius of curvature $l$ of the local magnetic field is of
the order of the crust thickness of the star (i.e. $l\sim 10^5$
cm), which is much less than the dipole radius of curvature $s$ of
dipole field component near stellar surface. The relation between
the local multipole magnetic field and dipole field can be given
by (Zhang \& Cheng 2003) \beq B_s\simeq B_d ({R\over l})^3. \enq
$B_d$ is the dipole magnetic field of a pulsar. So the critical
energy is rewritten as \beq E_{\rm crit}= {2m_ec^2\over
15}{B_q\over B_d}({R\over l})^{-3}. \enq

The secondary $e^\pm$ pairs will lose their energy via synchrotron
radiation with a photon energy at a distance of $r$ \beq E_{\rm
syn} ={3\over 2}({E_{\rm cur}\over 2m_ec^2})^2{e\hbar B(r)\over
m_ec}=2.6\times 10^5B_{d,9}P({3R\over r})^3({R\over l})^3 {\rm
eV}. \enq We can see that generally $E_{\rm syn}>E_{\rm crit}$, so
a photon-electron cascade will start and develop until this
condition fails. At the end of a cascade, each incoming primary
electron-positron can produce, on average \beq N_{e^\pm}={E_p\over
E_{\rm crit}}=1.9\times 10^3B_{d,9}P^{1/3}({R\over l})^3, \enq
then the total pair production rate can be estimated as \beq
\dot{N}_{e^\pm}=f\dot{N}N_{e^\pm}=2\times
10^{33}fB_{d,9}^{10/7}P^{-8/21}({R\over l})^{30/7}{\rm s^{-1}},
\enq where \beq \dot{N}=2.7\times 10^{27}P^{-2}B_{d,9}({R\over
l})^3 {\rm s^{-1}} \enq is the primary electron-positrons passing
through the polar gap (Goldreich \& Julian 1969).

Since these pairs are created close to the stellar surface and the
field lines are converging, only a small fraction may keep moving
toward the star and annihilate on the stellar surface. Ho (1986)
showed that the loss cone for these pairs will approach $\pi/2$
(cf. Eq. 29 of Ho 1986). In other words, most pairs will be
reflected by magnetic mirroring effect and then move toward the
light cylinder. These particles will flow out with the pulsar wind
and be accelerated by the low-frequency electro-magnetic wave. In
theory, the maximum Lorentz factor of pulsar wind particles could
be larger than $10^4$ (Rees \& Gunn 1974; Kennel \& Coroniti
1984).

\section{Three possible scenarios for pulsar winds in GC}
As discussed in the introduction, we consider three possibilities
of positron sources from pulsar winds in the Galactic center
region. In this section, we discuss them separately in detail.

\subsection{Normal pulsars}
The deep radio survey (LaRosa et al. 2000) and recent observations
in the Galactic center field targeted in the deep Chandra X-ray
surveys (Wang et al. 2002; Muno et al. 2003) showed that a few
supernova remnants are present. Detailed analyses of X-ray nebulae
G0.13-0.11, G359.89-0.08 and G359.54+0.18 (Wang, Lu \& Lang 2002;
Lu, Wang \& Lang 2003) suggested young pulsars might lie inside
them. So in the past years, young pulsars may have been produced
actively in the Galactic center region. However, because the
average proper motion velocity of normal pulsars is high ($\sim
500$ km s$^{-1}$), few of them ($\leq 40\%$) lie in the GC
(Arzoumanian, Chernoff, \& Cordes 2002), so the number of normal
pulsars whose active timescale is $\sim 1$ Myr in the GC should be
small, and the number of Crab or Vela like pulsars is around 10
(Wang, Jiang \& Cheng 2005).

There is both theoretical and experimental evidence that young
pulsars can emit intense electron/positron winds. In case of Crab
and Vela, the $e^\pm$ injection rates are up to $\sim 10^{41}$
e$^+$ s$^{-1}$ and $\sim 10^{39}$ e$^+$ s$^{-1}$. Then in the
timescale of the age of Crab or Vela, the total number of
positrons is about $3\times 10^{51}$. Assume the birth rate of
young pulsars is 1 per $10^3$ years in the GC, and the lifetime of
positrons is 1 million years (cf. the cooling time scale estimate
in section 4), there will be about $3\times 10^{54}$ positrons in
the region. On average the annihilation rate is $\sim 3\times
10^{54}/1 {\rm million\ yr} \sim 10^{41}{\rm s^{-1}}$. Therefore,
our analysis suggests that the young pulsar winds could not be a
significant contributor to the positrons in the Galactic center
region.

\subsection{Millisecond Magnetar in a GRB progenitor}
It has been suggested that gamma-ray bursts in the GC are
potential positron sources (Cass\'e et al. 2004; Parizot et al.
2005). Their models assumed that the annihilation line may relate
to nucleosynthesis in the hypernova which is the GRB progenitor.
But the GRB progenitor could also be a rapidly spinning strongly
magnetized neutron star (millisecond magnetar, see Usov 1992;
Thompson 1994), which is also a very strong $e^\pm$ pair emitter.
The generic magnetized $e^\pm$ outflow from a spinning compact
object could be given by \beq L_m\simeq 2\pi r_0^2 c({B^2\over
4\pi})\sim 1.5\times 10^{52}r_{06}^2B_{15}^2(t/t_0)^{-\alpha} {\rm
erg\ s^{-1}}, \enq $r_0$ is the size of the central engine, $\sim
10^6$ cm, $B\sim 10^{15}$ G is the magnetic field of magnetars,
$t_0$ is a characteristic timescale after the burst, i.e. $t_0\sim
1$ sec, where we have assumed the magnetic field decay would
follow a power law and $\alpha>1$. Assuming the average energy of
positrons is $\sim 1$ GeV, the total number of positrons emitted
by magnetars is \beq N\sim \int_{t_0}^{\infty} 2\times
10^{55}(t/t_0)^{-\alpha}dt\sim 2\times 10^{55}/(\alpha-1). \enq If
we take $\alpha=1.1$, the total $e^\pm$ number is up to $2\times
10^{56}$. For other possible values of $\alpha$, the total $e^\pm$
number will be reduced by a factor of several. On the other hand,
the toroidal magnetic field of magnetars can be up $10^{17}$G, and
part of this toroidal magnetic field energy can also be converted
into $e^\pm$ (Kluzniak \& Ruderman 1998; Dai \& Lu 1998).

Therefore, in this model more than $10^{56}$ positrons would be
produced. Because the GRB birth rate from the present observations
is $\sim 10^{-5}$ yr$^{-1}$ in a galaxy (Piran 2004) after
considering the beaming effect correction, we expect the GRB rate
in the GC is about one per $10^6$ years. Therefore, the
annihilation rate of positrons produced by the magnetar is $\sim
3\times 10^{42} {\rm s^{-1}}$. But more comments on the GRB rates
are required here. The observations of the long GRBs and host
galaxies are in agreement with a correlation of GRB locations with
blue regions/spirals of host galaxies (Bloom et al. 2002),
implying that long bursts do not concentrate in the bulge region;
the theoretical studies of merging neutron stars and black holes
suggest that most short bursts may be located in the halo region
(e.g Voss \& Tauris 2003) , so here we possibly have overestimated
the GRB rate in the Galactic center region considering the
uncertainties of burst sites of the different GRB populations.

\subsection{A millisecond pulsar population}
We have argued that millisecond pulsars could be the major
contributor to the pulsar population in the Galactic center (see
Wang, Jiang \& Cheng 2005). There may exist a population of old
neutron stars with low space velocities which have not escaped the
Galactic center (see Belczynski \& Taam 2004a; 2004b). Such
neutron stars could be recycled to millisecond periods, and remain
in the GC through their lifetime because of their relatively low
proper motion velocity (Lyne et al. 1998; Arzoumanian, Chernoff,
\& Cordes 2002). Wang , Jiang \& Cheng (2005) suggest that the
millisecond pulsar population can contribute significantly to the
diffuse gamma-rays in the Galactic center (Mayer-Hasselwander et
al. 1998). In addition, the X-ray synchrotron nebula around the
millisecond pulsar B1957+20 (Stappers et al 2003) provides
evidence that millisecond pulsars can remain active to emit
relativistic winds. Therefore, we propose that electron-positrons
from the winds of the millisecond pulsar population could be the
possible positron sources in the GC.

Since millisecond pulsars can lie in the GC and remain active for
1 Gyr, we believe they are continuous positron injecting sources.
For the typical parameters of millisecond pulsars, $P=3$ ms,
$B_d\sim 3\times 10^8$ G and according to equation 8, we estimate
the positron injection rate $\dot{N}_{e^\pm}\sim 5\times
10^{37}{\rm s^{-1}}$ for a millisecond pulsar. Wang et al. (2005)
showed that about 6000 millisecond pulsars could contribute to the
diffuse gamma-rays in the Galactic center region covered by EGRET.
Since the present annihilation observations show the line emission
region toward GC is about 6$^\circ$, is much larger than the EGRET
observation region (1.5$^\circ$), so a larger number of MSPs are
expected to lie in the bulge region of 6$^\circ$. We do not know
the distribution of MSPs in GC, so we just scale the number of
MSPs by $6000\times (6^\circ/1.5^\circ)^2\sim 10^5$, where we
assume the number density of MSPs may be distributed as
$\rho_{MSP}\propto r_c^{-1}$, where $r_c$ is the scaling size of
GC. Based on the population analysis of Lyne et al. (1998), the
number of millisecond pulsars in the entire Galaxy may exceed
$3\times 10^5$, so the number of millisecond pulsars in the bulge
region estimated here could be reasonable. So the total positron
injection rate from the millisecond pulsar population is $\sim
5\times 10^{42}$ e$^+$ s$^{-1}$.

\section{Electron-positron cooling in GC and annihilations}
The outflowing electron-positrons could be trapped by the magnetic
field in the Galactic center. The Larmor radius ($r_L$) of a
relativistic electron with energy $E_e$ is given by \beq
r_L\approx E_e/eB\sim 10^{13} (E_e/10^{10}{\rm eV})(B/10^{-5}{\rm
G})^{-1}{\rm cm}. \enq The angular size of $e^\pm$ annihilation
line emission detected by SPI/INTEGRAL is several degrees
(Kn\"odlseder et al. 2003), so we take $\lambda\sim 300$ pc as the
size of the Galactic center region we studied in this paper (we
assume the distance of the GC to us $\sim 8$ kpc). The diffuse
timescale of the electron is estimated as \beq t_{diff} \sim
({\lambda\over r_L})^2{r_L\over c}\sim 10^{10}(E_e/10^{10}{\rm
eV})^{-1}{\rm yr}, \enq where we have taken the average magnetic
field in the GC $B\sim 10^{-5}$ G (Uchida \& G\"usten 1995; LaRosa
et al. 2005) and $\lambda\sim 10^{21}$ cm.

Electrons and positrons in the Galactic center will lose their
energy via synchrotron radiation and Coulomb interactions with the
medium ,and then annihilate into 511 keV photons. If we assume the
injection electron-positron energy spectrum as the following form:
$\dot{Q}_{e^\pm}\sim A E_e^{-p}, $ where $p \geq 2$, then
positrons with energies higher than $\sim 100$ MeV will cool
through synchrotron radiation in the magnetic field, while a large
number of low energy positrons lower than $\sim 100$ MeV lose
their energy by Coulomb interactions in the medium of the GC. The
other energy loss processes, like inverse Compton cooling and
bremsstrahlung are negligible. When the positrons reach an energy
around a few tens of eV, they could either annihilate directly
with electrons or form positronium by charge exchange or radiative
capture (Bussard, Ramaty \& Drachman 1979). So the annihilation
timescales depend on the cooling timescales.

The synchrotron cooling timescale of relativistic positrons is
estimated as \beq \tau_s\simeq {\gamma m_ec^2\over
\gamma^2({eB\over m_ec})^2{e^2\over c}} \sim 3\times 10^{6}
(E_e/10^{10}{\rm eV})^{-1} B_{-5}^{-3/2}{\rm yr}. \enq We find
that the electron synchrotron timescale is lower than the
diffusion timescale for the same electron energy, $\tau_s \ll
t_{diff}$, so the positrons could significantly lose their energy
to become non-relativistic, and stay in the Galactic center.

The energy loss timescale of Coulomb interactions in the medium,
$\tau_c$ depends on the positron energy, the medium density and
the degree of ionization. For the positrons with energy around
1-100 MeV, the Coulomb interaction cooling timescale is estimated
as $\tau_c \sim 10^5 n^{-1}$ yr (within a factor of 2), and $n$ is
the medium number density in units of cm$^{-3}$. However, the
medium in GC is quite complicated, there exist many giant
molecular clouds (LaRosa et al. 2000), and hot gas discovered in
X-rays (Muno et al. 2004). Then the energy loss timescales could
vary from $\tau_c<10^3$ yr in the molecular clouds ($n>10^2{\rm
cm^{-2}}$), $\tau_c\sim10^5$ yr for the typical warm interstellar
medium ($n\sim 1{\rm cm^{-2}}$), and $\tau_c>10^7$ yr for the hot
gas ($n\sim 10^{-2}{\rm cm^{-2}}$). The analysis of the 511 keV
annihilation spectrum suggests that the dominant fraction of
positrons ($\sim 94\%$) form positronium before annihilation,
which constrains the positron annihilation in the hot gas to a
very small fraction ($<8\%$, see Churazov et al. 2005). Most
positron annihilations seem to lie in the warm medium, so the
cooling timescale is also much lower than the diffusion timescale.

\section{Summary and discussions}

The origin of Galactic positrons is still a mystery at present. In
this paper, we have proposed a possible contribution to positron
sources from the pulsar winds in the Galactic center region. We
discussed the contributions of three possible pulsar scenarios in
the GC: normal pulsars; the rapidly spinning strongly magnetized
neutron stars in GRB progenitors (millisecond magnetars); a
millisecond pulsar population. Electron-positron pairs are
injected into the Galactic center region from three classes of
pulsar winds, and could be trapped in this region by the magnetic
field in the Galactic center. These relativistic $e^\pm$ pairs
will lose their energy though the synchrotron radiation and the
Coulomb interactions with the medium to become non-relativistic.
The cooling timescales are lower than the diffusion timescale of
positrons in the magnetic field in the GC. So positrons could form
positronium by charge exchange or radiative capture, or annihilate
directly with electrons into photons to produce the 511 keV line
observed by the present detectors. Our results have shown that
normal pulsars cannot be a significant contributor to positron
sources in GC, but the millisecond magnetar and a millisecond
pulsar population could be potential positron sources. However, as
we have estimated that the diffusion time is much longer than the
cooling time, it is not clear how the positrons produced by GRBs
can spread over the entire galactic bulge or at least in the range
of 6$^{\circ}$. Parizot et al. (2005) have argued that the
turbulent diffusion process is able to diffuse the positrons over
the galactic bulge in a time scale of $10^7$yr. However, it is
still not clear why the positrons produced by GRBs should
concentrate in the galactic bulge, which is not supported by the
observations of GRBs and their host galaxies (e.g. long bursts in
Bloom et al. 2002; short bursts in Gehrels et al. 2005). In view
of the uncertainty of the birth rate of GRBs in the GC and lacking
of annihilation lines in disk, we have some reservations on GRBs
as the significant positron sources in the GC.

The predicted annihilation rate of positrons from the millisecond
pulsar population is $\sim 5\times 10^{42}$ s$^{-1}$ which is
consistent with the present observational constraints by
SPI/INTEGRAL. So we conclude that the $e^\pm$ pairs from these
classes of pulsar winds are potential positron sources, and can
significantly contribute to the 511 keV annihilation lines in GC.

The important problem for positrons is to determine the $e^+e^-$
annihilation line intensity and radiation morphology in the
Galaxy. Recently, analyses of the data by SPI/INTEGRAL with deep
Galactic center region exposure show that the spatial distribution
of 511 keV line appears centered on the Galactic center (bulge
component), with no contribution from a disk component (Teegarden
et al. 2005; Kn\"odlseder et al. 2005; Churazov et al. 2005). We
suggested that the positrons produced by pulsar winds could be
trapped in the Galactic center by the magnetic field, the
diffusion timescale is higher than the cooling timescales, so that
the positrons can annihilate into 511 keV photons before they can
escape from the Galactic center region. In the other viewpoints,
the bulge dominated morphology for 511 keV line may indicate that
the positron source population could be an old stellar population.
Kn\"odlseder et al. (2005) have suggested that the low-mass X-ray
binaries (LMXBs) could be the candidate sources because more than
60\% of the Galactic LMXBs are observed towards the galactic bulge
(see Grimm et al. 2002). The millisecond pulsar population is a
very old stellar population, and part of them stay in LMXBs, but
most may be isolated as radio pulsars, X-ray sources, even
detectable in gamma-rays. The present radio observations find most
millisecond pulsars lie in globular clusters (see the review by
Camilo \& Rasio 2005 and the references therein), few of them in
the Galactic field. If assuming the millisecond pulsar
evolutionary formation channels in globular clusters are similar
to the Galactic center region or bulge region, then we expect a
millisecond pulsar population in the GC as a significant
contribution to the positron sources. The millisecond pulsars can
stay in the Galactic center region throughout their lifetime,
which is well consistent with the 511 line emission morphology.

We have shown that the millisecond pulsar population in the
Galactic center could be the major sources of positrons. Normal
pulsars could not be the main positron sources, but GRB
progenitors,i.e. millisecond magnetars, may be significant
contributors. In addition hypernovae/GRBs could also be the
potential positron sources (Cass\'e et al. 2004). Thus, how could
we discriminate the model of a millisecond pulsar population from
other models, specially models related to GRBs? Firstly, we can
estimate the typical spatial diffusion scale, according to Eq.
(13), $\lambda_{diff}\sim (r_L ct)^{1/2}$, the average cooling
time of positrons in GC is $10^6$ years (Churazov et al. 2005), so
the characteristic diffusion scale is about $10^{18}$ cm. Because
of the low angular resolution of SPI/INTEGRAL (about 2 degrees,
Vedrenne et al. 2003), we can assume that the positrons annihilate
in the same local region as their sources, i.e. the millisecond
pulsars. Therefore we predict that the spatial intensity
distribution of the annihilation lines should follow the spatial
distribution of MSPs if there exists a millisecond population in
the GC. Additionally, we could assume the spatial distribution of
MSPs should follow the mass distribution of the GC though we do
not know how well they follow each other. But because the proper
motion velocity of MSPs is relatively low, we could reasonably
assume two distributions are quite close to each other. Therefore,
if the positron sources originate from the MSP population, the 511
keV annihilation line intensity would follow the mass distribution
of the Galactic center region. The present image of 511 keV line
emission (see Figs. 4 and 6 in Kn\"odlseder et al. 2005) shows
that the flux is strongest in the center, and decreases toward the
outside region, possibly tracing the mass distribution of the GC,
but this needs confirmation by high resolution observations in
future. The millisecond pulsar population could naturally explain
the diffuse emission morphology and does not have the problem of
the turbulent diffusion, which is required to diffuse all these
positrons to a few hundred pc. On the other hand, for models
associated with GRBs, the positron source is initially a
point-like source, so the diffuse emission of 511 keV line needs
the strong turbulent diffusion to diffuse the positrons to a
larger scale (Parizot et al. 2005). In the present viewpoints, GRB
progenitors (either magnetars or hypernovae) are related to
massive stars and dense clouds (e.g. molecular clouds in GC which
has a dimension of a few tens pc, LaRosa et al. 2000), the mean
density $n>10^2 {\rm cm^{-3}}$, the cooling time of positrons is
less than $10^3$ years. So the spatial diffusion scale is only
$10^{17}$ cm, which makes the annihilation line emission looks
like a point source correlated with the molecular clouds. In order
to make the line emission be a diffuse source, GRB positrons must
diffuse to a hundred pc with a long diffusion timescale $> 10^7$
years, so it requires that the mean density in the GC is $\sim
10^{-2}{\rm cm^{-3}}$. But this low density seems unreasonable in
the GC, and also is not consistent with the constraint on the
medium environment from the spectrum of 511 keV line (Churazov et
al. 2005). Furthermore, if positrons cannot diffuse to larger
distances, the intensity distribution of annihilation lines should
not correlate with the general mass distribution of the GC,
instead it should be more correlated to the distribution of
massive stars and dense molecular cloud. Moreover, GRBs more
likely occur in the disk, where more dense molecular clouds and
massive stars are located, and in fact the detected magnetars and
soft gamma-ray repeaters are also distributed in the disk. Giant
flares from soft gamma-ray repeaters (SGR) are also positron
source candidates which are distributed in the Galactic plane, but
the 511 keV line observations show a very weak disk component
(Teegarden et al. 2005; Kn\"odlseder et al. 2005). The 27 Dec.
giant flare from SGR 1806-20 released an energy up to $10^{47}$
erg (Hurley et al. 2005), and is thought to be a possible origin
of short GRBs, and could be observed as the 511 keV emission
source in future. In summary, the positron source models
associated with GRBs are disputable. Our scenario of a millisecond
pulsar population as possible positron sources in the GC has some
advantages to explain the diffuse morphology of 511 keV line
emissions, and predicts the line intensity distribution should
follow the mass distribution of the GC, which may be tested by
future high resolution observations.

\begin{acknowledgements}
We are grateful to the referee for the critical comments, to R.
Diehl, Y. F. Huang and A. Strong for the discussions. This work is
supported by a RGC grant of the Hong Kong Government under HKU
7015/05P and the National Natural Science Foundation of China
under grant 10273011.
\end{acknowledgements}

\end{document}